\documentclass[journal,twoside,web]{ieeecolor}

\usepackage{generic}

\usepackage{comment} % to comment out blocks of text
\usepackage{graphicx}
\usepackage{subcaption}
\usepackage{url}
\usepackage{amsfonts}
\usepackage{amsmath}
\usepackage{bm} 
\usepackage{xcolor}
\usepackage{arydshln}
\usepackage{mathtools}
\usepackage{cite}
\usepackage[colorlinks,pdftex,pdfpagelabels,bookmarks,hyperindex,hyperfigures]{hyperref}
\hypersetup{
  pdftitle={},
  pdfauthor={Reza Sameni},
    pdfnewwindow=true,      % links in new window
    colorlinks=true,       % false: boxed links; true: colored links
    linkcolor=blue,          % color of internal links
    citecolor=magenta,        % color of links to bibliography
    filecolor=cyan,      % color of file links
    urlcolor=teal%magenta           % color of external links		
}
\DeclareMathOperator{\Diag}{\mathbf{diag}}

\DeclareMathOperator*{\argmin}{arg\,min}

\graphicspath{ {./images/} }

\begin{document}
%//////////////////////////////////////////////////////////////////////////////////////
% \title{Robust T-wave Alternans Detection and Measurement???}

% \title{Noise Robust Methods for T-wave Alternans Detection and Measurement}

% \title{Robust T-wave Amplitude Measurement with Alternans Detection with Model-Based Techniques}

% \title{Robust Measurement of T-wave Amplitude and Alternans using Model-Based Techniques}

\title{A Noise-Robust Model-Based Approach to T-Wave Amplitude Measurement and Alternans Detection}

% T-wave measurement
% modeling
% TWA detection
% noise stressing

%//////////////////////////////////////////////////////////////////////////////////////////////
\author{Zuzana Koscova, Amit Shah, Ali Bahrami Rad, Qiao Li, Gari D Clifford, Reza Sameni %
% \\
% \textcolor{red}{Gari, Reza, Amit, please reorder the authors to whatever you think is most appropriate}
\thanks{Z.~Koscova, A.~Bahrami Rad, Q.~Li, G.D.~Clifford, and R.~Sameni are with the Department of Biomedical Informatics, Emory University School of Medicine. A.~Shah is with the Department of Epidemiology, Rollins School of Public Health, and the Department of Medicine (Cardiology), Emory University School of Medicine. G.D.~Clifford and R.~Sameni are also with the Department of Biomedical Engineering, Georgia Institute of Technology and Emory University.
 (email: \url{zuzana.koscova@dbmi.emory.edu}, \url{rsameni@dbmi.emory.edu}).
}%
%\thanks{Copyright (c) 2019 IEEE. Personal use of this material is permitted.  However, permission to use this material for any other purposes must be obtained from the IEEE by sending an email to pubs-permissions@ieee.org.}%
}%\newline
%//////////////////////////////////////////////////////////////////////////////////////////////
% \markboth{}{}
%\pubid{0000--0000/00\$00.00~\copyright~2007 IEEE}
% \date{\today}
\date{December 2024}
\maketitle

\begin{abstract}
T-wave alternans (TWA) is a potential marker for sudden cardiac death, but its reliable analysis is often constrained to noise-free environments, limiting its utility in real-world settings. We explore model-based T-wave estimation to mitigate the impact of noise on TWA level. Detection was performed using a previous surrogate-based method as a benchmark and a new method based on a Markov model state transition matrix (STM). These were combined with a Modified Moving Average (MMA) method and polynomial T-wave modeling to enhance noise robustness. Methods were tested across a wide range of signal-to-noise ratios (SNRs), from -5 to 30\,dB, and different noise types: baseline wander (BW), muscle artifacts (MA), electrode movement (EM), and respiratory modulation. Synthetic ECGs with known TWA levels were used: 0\,$\mu$V for TWA-free and 30–72\,$\mu$V for TWA-present signals.

T-wave modeling improved estimation accuracy under noisy conditions. With EM noise at SNRs of -5 and 5\,dB, mean absolute error (MAE) dropped from 62 to 49\,$\mu$V and 27 to 25\,$\mu$V, respectively (Mann–Whitney-U test, $p < 0.05$) with modeling applied. Similar improvements were seen with MA noise: MAE dropped from 100 to 70\,$\mu$V and 26 to 23\,$\mu$V.

In detection, the STM method achieved an F1-score of 0.92, outperforming the surrogate-based method (F1 = 0.81), though both struggled under EM noise at -5\,dB. Importantly, beyond SNR, detection performance depended on the number of beats analyzed.

These findings show that combining model-based estimation with STM detection significantly improves TWA analysis under noise, supporting its application in ambulatory and wearable ECG monitoring.

\end{abstract}

\section{Introduction}
\label{sec:introduction}
T-wave alternans (TWA) is characterized by beat-to-beat alternations in the amplitude or shape of the T-wave on an electrocardiogram (ECG). It has been reported as a critical marker for cardiac risk stratification, particularly for conditions such as sudden cardiac death \cite{chow2007usefulness, bloomfield2006microvolt}. However, its accurate measurement and detection are significantly affected by the inherent variability of T-wave amplitude, which is highly susceptible to measurement of noise and respiratory effects. Specifically, respiration effects tend to modulate the ECG amplitude rather than act as additive noise, limiting the effectiveness of conventional filtering techniques. These challenges are further compounded by the limited availability of robust methods for TWA analysis and the absence of a standardized dataset for benchmarking.

Currently, the most common approach for T-wave measurement relies on a modified moving average (MMA) method \cite{nearing2002mma}, which has demonstrated strong performance in controlled settings. However, this method has not been stress-tested under additive noise and respiratory modulation. Similarly, the surrogate data analysis (SDA), which is a well-established method for TWA detection \cite{nematitwa}, enables the differentiation of true alternans from noise. Despite these advances, the performance of these methods remains largely unexplored in the presence of realistic noise levels, particularly in data collected from ambulatory and wearable ECG devices, where noise is both prevalent and compound. 

The distinction between TWA ``measurement'' and ``detection' is critical. Measurement quantifies the magnitude of TWA, whereas detection determines its presence. Both aspects are essential, but from an estimation-theoretical perspective, they are distinct problems requiring optimization through different frameworks. In noisy environments, the reliability of both measurement and detection diminishes, highlighting the need for novel objective methods that incorporate noise resilience to improve both TWA measurement and detection.

Model-based approaches offer a promising solution, as they introduce priors and assumptions that can enhance algorithmic robustness to noise. In this paper, we propose an objective model-based framework that advances both T-wave measurement and TWA detection. We develop a novel approach that integrates ECG waveform modeling with existing TWA assessment methods, demonstrating enhanced accuracy in measurement and sensitivity in detection. A new detection method, based on the state transition matrix (STM) of consecutive T-wave amplitudes, is introduced to model TWA and distinguish true alternans from random T-wave fluctuations caused by noise or respiratory effects. This method is shown to be combinable with the Modified Moving Average technique and the surrogate data analysis, providing a hybrid framework for comprehensive TWA assessment. To ensure robustness, the proposed methods are rigorously evaluated under varying signal-to-noise ratios (SNR) and across different noise types, including respiration-modulated noise.

Given the lack of standardized TWA datasets, we approach the problem by generating a synthetic database of ECG signals with controllable TWA and noise characteristics. Using a well-established synthetic yet realistic ECG generation model, we simulate waveforms resembling ECG records (which we train on the PTB dataset \cite{Bousseljot1995}, with controlled heart rates, T-wave amplitudes, respiratory effects, and noise. These waveforms mimic real ECG records and provide ground truth T-wave amplitudes, which serve as references for evaluating TWA measurement and detection algorithms. This enables systematic testing and benchmarking. While the proposed methods are validated through synthetic experiments, further clinical validation will be necessary to confirm their practical utility. This research lays the groundwork for improving TWA measurement and detection, particularly in the increasingly noisy environments of wearable and ambulatory ECG monitoring.

%/////////////////////////////////////////
\section{Literature review}
T-wave alternans (TWA), characterized by beat-to-beat variations in the timing or morphology of the ST-T complex on the surface ECG, was first identified by Hering in 1908~\cite{Hering1908}. In 1981, Adam et al. discovered microvolt-level TWA, which is too subtle to be detected visually on standard ECG displays~\cite{Adam1981}. Today, TWA is recognized as a significant marker for sudden cardiac death~\cite{Armoundas2002, takagi2003twave, bloomfield2004microvolt, CUTLER2009S22}.

Various methods have been developed for TWA estimation, incorporating both time-domain and frequency-domain techniques to detect and analyze alternans in ECG signals. For a comprehensive and systematic overview of these methods, see the review by Mart\'inez and Olmos~\cite{Martinez2005}. Former approaches for TWA analysis include the Energy Spectral Method~\cite{Adam1981, Adam1982, Adam1984}, the Spectral Method~\cite{Smith1988}, the Complex Demodulation Method~\cite{Nearing1991}, the Correlation Method~\cite{Burattini1997}, the Karhunen-Lo\`eve Transform~\cite{Laguna1996}, the Capon Filtering Method~\cite{Martinez2000}, the Poincar\'e Mapping Method~\cite{Strumillo2002}, the Periodicity Transform Method~\cite{Srikanth2002}, the Statistical Tests Method~\cite{Srikanth2002_2}, the Laplacian Likelihood Ratio Method~\cite{Martinez2002}, among others.

Currently, the state-of-the-art method for TWA estimation from ECG is the modified moving average (MMA) approach, introduced by Nearing et al. in 2002 \cite{nearing2002mma}. This time-domain method computes separate moving averages for even and odd beats. Risk stratification is based on the peak TWA level, with thresholds of 47\,$\mu$V and 60\,$\mu$V indicating abnormal and severely abnormal levels, respectively.

A notable limitation in TWA research is the absence of a human-annotated TWA database. During the PhysioNet Challenge 2008 \cite{moody2008physionet}, participants were tasked with ranking synthetic and real ECG recordings based on TWA levels. Despite employing a variety of techniques, none outperformed the MMA method. Existing approaches require noise-free ECGs, and recordings with poor signal quality or high noise levels are often misclassified or excluded from analysis.

Sadiq et al. (2021) \cite{sadiq2021breathing} showed that physiological breathing rate (BR) and heart rate (HR) significantly influence TWA, highlighting the importance of accounting for BR and HR as independent confounding variables.

To address noise-induced TWA misestimation, Nemati et al. \cite{nematitwa} applied a nonparametric surrogate-based test to assess the significance of TWA detected using the MMA method.

In this study, we aim to evaluate TWA under a range of conditions, including various noise types, differing signal-to-noise ratios (SNRs), and the presence or absence of respiratory modulation. We will apply established methods such as MMA and the spectral domain approach (SDA), along with novel techniques including T-wave modeling and classification of TWA based on a Markov model of T-wave amplitudes. By integrating these methods, we aim to improve TWA estimation and detection, particularly in recordings with low SNR.

%%\textcolor{blue}{Zuzana, please refer to the TWA literature using the PhysioNet Challenge. Also highlight the surrogate data analysis and modified moving average approach.
%}

%%Cite this: 
%\url{https://iopscience.iop.org/article/10.1088/1361-6579/abd237/}

%/////////////////////////////////////////
\section{Method}
\subsection{Synthetic ECG generation}
\label{sec:dipole_model}

To objectively evaluate the performance of various T-wave amplitude measurement and TWA detection methods under different noise levels and respiratory effects, we use a well-established synthetic ECG generation model capable of producing realistic ECG waveforms with controlled noise and artifact levels. An overview of this model and the training of its parameters on real ECG datasets is provided below.

% \subsection{A dipolar model of the ECG}
Various models have been proposed for body surface ECG~\cite{MP95}. According to the dipolar model of the heart \cite[Ch.~15]{MP95}, signals from body surface leads can be modeled as projections of a single (equivalent) cardiac dipole vector onto electrode axes. Due to the properties of the body \textit{volume conductor}, these signals are quasi-periodically synchronous with the cardiac phase. These properties have been used to create synthetic models for body surface cardiac waveforms~\cite{McSharry2003,cliffordSPIE04,GDC06}, which have been extended to multilead vectorcardiogram (VCG) models for adult and fetal ECGs \cite{SSJB05,SCJS06}. In \cite{SCJS06}, using the single dipole model of the heart and a generalization of the McSharry-Clifford model \cite{McSharry2003}, the following dynamic model was proposed for simulating the three dipole coordinates of the VCG, denoted as $\mathbf{s}(t)=[x(t), y(t), z(t)]^T$:
\begin{equation}
\begin{array}{l}
\displaystyle\dot{\theta}=\omega\\
\displaystyle\dot{x}=-\sum_{i}\frac{\alpha^x_i\omega\Delta\theta^x_i }{(b^x_i)^2}\exp[-\frac{(\Delta\theta^x_i)^2}{2(b^x_i)^2}]\\
\displaystyle\dot{y}=-\sum_{i}\frac{\alpha^y_i\omega\Delta\theta^y_i}{(b^y_i)^2}\exp[-\frac{(\Delta\theta^y_i)^2}{2(b^y_i)^2}]\\
\displaystyle\dot{z}=-\sum_{i}\frac{\alpha^z_i\omega\Delta\theta^z_i}{(b^z_i)^2}\exp[-\frac{(\Delta\theta^z_i)^2}{2(b^z_i)^2}]
\end{array}
\label{eq:SyntheticDipole}
\end{equation}
where $\Delta\theta^x_i=(\theta-\theta^x_i)\mod(2\pi)$, $\Delta\theta^y_i=(\theta-\theta^y_i)\mod(2\pi)$, $\Delta\theta^z_i=(\theta-\theta^z_i)\mod(2\pi)$, and $\omega=2\pi f$ is the cardiac angular velocity, with $f$ being the instantaneous heart-rate in Hz. The first equation in \eqref{eq:SyntheticDipole} generates a circular trajectory corresponding to the cardiac cycle, while the other equations describe the dynamics of the dipole coordinates in the 3D space, as a sum of Gaussian functions with varying amplitudes, widths, and phase. These Gaussians adjust the dipole vector trajectory in the $(x,y,z)$ space, creating a moving vector. This rotating dipole model is very general, as, according to the universal approximation theorem, a sum of Gaussian mixtures can approximate any continuous function with arbitrary accuracy~\cite{bb16291}.

The electrophysiology of the ECG has three properties: 1) \textit{quasi-static electromagnetic behavior}: electric and magnetic fields are decoupled, with the electric field proportional to the gradient of the scalar potential, and the current density having zero divergence; 2) \textit{linear superposition}: the electrical potentials from the heart and other biopotential sources superimpose linearly; and 3) \textit{resistive tissue impedance}: at low frequencies (below 10\,kHz for the ECG), body tissues exhibit predominantly resistive electrical impedance, with negligible capacitive effects. As a result, the cardiac dipole can be mapped to body surface ECG leads using a linear transform:
\begin{equation}
    \mathbf{x}(t) = \mathbf{H} \mathbf{s}(t) + \bm{\alpha}\mathbf{n}(t)
\label{eq:data_model}    
\end{equation}
where $\mathbf{x}(t)\in\mathbb{R}^{n}$ represents body surface ECG leads, $\mathbf{H}(t)\in\mathbb{R}^{n\times 3}$ represents the volume conductor matrix, $\mathbf{s}(t)\in\mathbb{R}^{3}$ is the cardiac diapole, $\mathbf{n}(t)\in\mathbb{R}^{n}$ represents the per-lead measurement noise, and $\bm{\alpha}=\Diag{(\alpha_1, \ldots, \alpha_n)}$ is a diagonal matrix, where $\alpha_k$ ($k = 1, \ldots, n$) serve as scaling factors for controlling the channel-wise noise levels as detailed in Section~\ref{sec:noise_gen}.

In practice, the three orthogonal Frank leads, which are used for a vectorcardiogram (VCG) representation of the ECG, can be used to model $\mathbf{s}(t)$, and the twelve (or fewer) leads can be considered as $\mathbf{x}(t)$~\cite{SCJS06}. Then, for datasets like the Physikalisch-Technische Bundesanstalt database (PTBDB)~\cite{Bousseljot1995}, which include both the 12-lead ECG and the three VCG leads, $\mathbf{H}$ can be obtained by solving a least squares problem:
\begin{equation}
 \mathbf{H}^* = \argmin_H \| \mathbf{x}(t) - \mathbf{H} \mathbf{s}(t) \|   
\end{equation}
which is essentially a generalization of the Dower transform~\cite{Dower1980}.

By introducing randomness to the model parameters and introducing noise, more realistic ECGs with inter-beat variations can be simulated~\cite{SCJS06}. Modulatory effects such as respiration can be introduced via the volume conduction matrix $\mathbf{H}$, making it a function of time~\cite{SCJS06}. The model has been shown to be well-suited for ECG waveforms of arbitrary morphology. It has also been coupled with Markov models to generate combinations of normal and ectopic beats, as well as alternating behaviors observed in TWA~\cite{Clifford2010}.

\subsection{Training the synthetic model parameters from real ECG}
The ECG model introduced in Section~\ref{sec:dipole_model} is parametric, and its parameters can be selected to mimic specific ECG recordings. We used PTBDB~\cite{Bousseljot1995}, to create our dataset of synthetic ECGs with varying amplitudes of TWA.

The initial morphologies of these artificial ECGs were derived using a least-squares fit of Gaussian parameters on the three VCG Frank leads from 20 normal subjects in PTBDB. For this, the R-peaks of one-minute segments from each record were detected. Next, the average beats for each record were estimated from the VCG leads (V\textsubscript{x}, V\textsubscript{y}, and V\textsubscript{z}). Gaussian functions were then fitted to the average beats using the interactive \texttt{ecg\_beat\_fitter\_gmm\_gui} tool from the Open-Source Electrophysiological Toolbox (OSET)~\cite{Sameni_OSET,ismail2021}, which fits a sum of Gaussian waves to the average beat using a nonlinear least-squares approach~\cite{SCJS06}. The number of Gaussians was fixed to 11 for all records, and the Gaussian kernel positions were selected such that the T-wave segment would be modeled with exactly two Gaussian functions. This resulted in a good modeling fit (below 5\% error vs. the average beat for all records) and simplified subsequent model fitting by maintaining an identical number of Gaussian kernels over the T-wave segment.

% from  best approximate the average beat morphology in each VCG recording by minimizing the squared error \cite{ismail2021}. The Dower transform was then applied to the artificial VCG to generate the twelve-lead ECG representation \cite{Dower1980}. 

% The intuition behind this approach was that any continuous function can be approximated arbitrarily well with a finite number of Gaussian functions.

Even though different leads and abnormal beats may exhibit completely different morphologies compared to normal beats or may have minor variations in specific regions depending on the type of abnormality, under the \textit{universal approximation theorem} for the sum of Gaussian functions (Radial Basis Function approximation)~\cite{Park1991}, a sufficient number of Gaussian functions can be used to precisely replicate the average beat to synthesize both normal and abnormal beats~\cite{SCJS06,CNS2010}.

After obtaining the parameters of the Gaussian kernels representing the average ECG waveforms of the twenty PTBDB subjects, synthetic ECG signals were generated at four distinct heart rates (HRs): 60, 70, 80, and 90 beats per minute. Respiration was modeled as:
\begin{equation}
    r(t) = 1 + a \sin(2\pi f_b t + \phi_0)
\end{equation}
where $f_r = \text{RR}/60$ and $\text{RR}$ (respiration rate) was selected between four values: 12, 16, 18, and 24 breaths per minute. $a$ is the RR amplitude fluctuation, set to 0.1 in the following results (equivalent to 10\% of respiratory-induced ECG amplitude fluctuations), and $\phi_0$ was set randomly for each instance of the RR signal. The resulting $r(t)$ was multiplied by the leads of $\mathbf{x}(t)$ to model the non-additive modulatory effect of respiration on the ECG amplitude.

To model the TWA effect, TWA was introduced by modifying the T-wave amplitude in every second beat, by adding an offset to the 10th and 11th Gaussian kernels. Offset values ranged from 30 to 72\,$\mu$V in 7\,$\mu$V increments. This was inspired by previous research on cut-points of 47\,$\mu$V and 60\,$\mu$V as thresholds for abnormality and severe abnormality, respectively \cite{Verrier2011, Lewek2017}. By including values slightly below and above these thresholds, we aimed to assess the detectability of lower amplitudes and to enrich our dataset. Datasets without TWA were generated without altered T-wave amplitudes. 

All signals were created at a sampling rate of 1\,kHz with a duration of 60 seconds. For further analysis, we will focus solely on lead I, as this ensures the method remains applicable in home settings, where the environment is particularly noisy.

% For datasets with T-wave alternans (TWA), the TWA level ranged from 47 to 89 µV for each signal. These levels were selected according to guidelindefined cut-points of 47 µV and 60 µV, which represent thresholds for abnormality and severe abnormality, respectively \cite{Verrier2011, Lewek2017}. Conversely, datasets without TWA were generated without any beats exhibiting altered T-wave amplitudes.

% In this study, T-wave alternations (TWA) were generated by modifying the amplitude of the T-wave in every second beat. Normal beats were modeled using 11 Gaussian functions, while abnormal beats (alternate beats in the TWA time series) were generated by adding an offset to the amplitudes of the 10th and 11th Gaussians corresponding to the T-wave \cite{Clifford2010}. The offset values were arbitrarily chosen between 47 and 89 $\mu$V, with increments of 7 µV.
% Since TWA activity predominantly occurs in a preferred plane and only a single lead (lead I) is used in this study for TWA estimation, the exact TWA level in lead I was determined as follows: only Gaussian functions corresponding to the T-wave (10th and 11th) were used to generate the ECG signal for both normal (even beats without T-wave abnormalities) and abnormal (odd beats with T-wave abnormalities) beats. The peak amplitudes of the T-waves were detected, and the difference between these amplitudes for normal and abnormal beats was calculated to quantify the exact level of TWA in lead I.

\subsection{Noise generator}
\label{sec:noise_gen}
After generating clean synthetic ECG signals with respiratory effects, TWA, and controlled heart rate, we added one of three noise types from the Noise Stress Test database \cite{moody1984noise,goldberger2000physiobank} to the synthetic ECG. The noise types considered were baseline wander (BW), muscle artifacts (MA), and electrode movements (EM). Each noise type was injected at different signal-to-noise ratio (SNR) levels, ranging from -5\,dB to 30\,dB in 5\,dB increments.

\begin{figure}[tb]
%\centering
\includegraphics[width=\columnwidth]{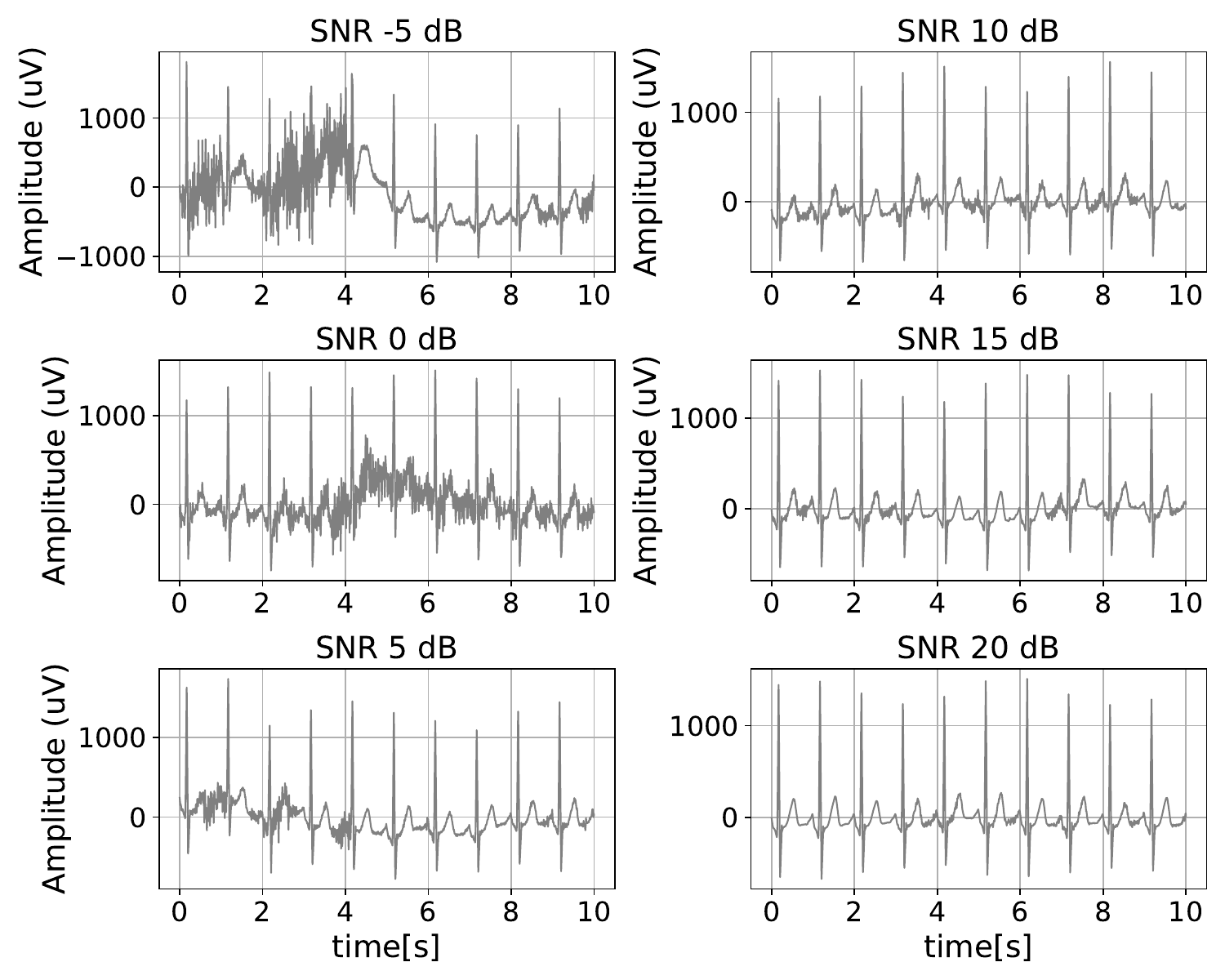}
\caption{Example of a signal with respiratory modulation and muscle artifact noise at varying SNR levels (-5\,dB to 20\,dB). 
%Each subplot shows a 10-second noise segment extracted from a 30-minute noise recording and added to the clean signal.
}
\label{fig:snr_example}
\end{figure}

The noise recordings were originally sampled at 360\,Hz and had a duration of 30 minutes. In order to match the sampling frequency and specified SNR levels, the noise samples were first resampled to 1\,kHz to match the sampling frequency of the synthetic ECG signals. A random continuous 60-second segment was then extracted to be added to the clean ECG. To achieve the desired SNR, the appropriate scaling factor $\bm{\alpha}$ in \eqref{eq:data_model} was calculated as follows:
\begin{equation}
    \alpha_k = \displaystyle\sqrt{\frac{\mathbb{E}_t[y_k^2(t)]} {\mathbb{E}_t[n_k^2(t)]}}10^{-\displaystyle\frac{\text{snr}}{20}}
\end{equation}
where $k$ corresponds to the $k$-th lead; $\alpha_k$ is the $k$-th element of the vector $\bm{\alpha}$; and $n_k(t)$ is the noise to be added to channel $k$ (before scaling by $\alpha_k$). Denoting $\mathbf{y}(t) := \mathbf{H}\mathbf{s}(t)$, $y_k(t)$ is the $k$-th element of $\mathbf{y}(t)$, and $\mathbb{E}_t(\cdot)$ denotes averaging over time. 
A typical example of 10\,s ECG segment at various levels of SNR for muscle artifacts noise is illustrated in Fig.~\ref{fig:snr_example}.

\subsection{Evaluation}
To assess the detection and estimation of T-wave alternans (TWA), we created four distinct datasets, each containing synthetic ECG with and without TWA. The datasets are as follows: (1) clean synthetic ECG, (2) ECG with respiratory modulation, (3) ECG with additive noise, and (4) ECG with both respiratory modulation and additive noise. All datasets consist of the same set of records, generated from identical set of Gaussian parameters. The number of records in each dataset is listed in Table~\ref{tab:dataset}.

\begin{table}[tb]
\centering
\caption{Types and number of synthetic ECG datasets generated for this study}
\begin{tabular}{c|c|c|c}
\hline
\textbf{Dataset} & \textbf{Description} & \textbf{\# TWA} & \textbf{\# TWA-free}\\
\hline
1 & Clean &  560 & 80 \\
\hline
2 & Respiratory modulated & 2,240 & 320 \\
\hline
3 & Noise corrupted & 13,440  &  1,920\\
\hline
4 & Respiratory modulated and noisy & 53,760 &  7,680\\
\hline
\end{tabular}
\label{tab:dataset}
\end{table}

%To evaluate the detection and estimation of T-wave alternans (TWA), we generated a dataset consisting of 47,040 recordings with varying levels of TWA and 6,720 recordings without TWA. Recordings included various types of noise with different SNR, HR, and RR. 

Our evaluations are across three aspects: (a) measuring T-wave amplitude, (b) estimating TWA using the Modified Moving Average~\cite{nearing2002mma}, and (c) classifying recordings as either TWA or TWA-free using two methods---the Markov model state transition matrix and surrogate data analysis.

The first step in all three methods is data pre-processing. We estimated baseline wander by applying a second-order, zero-phase, high-pass filter with a cutoff frequency of 0.1\,Hz. Following baseline correction, R-peaks were detected using the \texttt{peak\_det\_likelihood.m} function from the Open-Source Electrophysiological Toolbox (OSET)~\cite{Sameni_OSET}. The identified R-peaks were next processed using the \texttt{wavedet\_3D.m} function from the ECG-Kit toolbox~\cite{Demski2016} to extract fiducial points of the ECG, including the T-wave onset, offset, and peak. The subsequent steps of method are detailed below.

\section{T-wave amplitude analysis}
In the first step, we aimed to quantify the T-wave amplitude and compare it with the known ground truth. Since the synthetic data were generated using a mixture of Gaussian kernels, the exact T-wave amplitudes prior to the addition of respiratory modulation and various types and levels of noise are known. 

\subsection{T-wave modeling}
After detecting the T-wave peaks using \texttt{wavedet\_3D.m}, the amplitudes were calculated to construct the T-wave amplitude sequence for each recording. Apparently, accurate T-wave amplitude calculation becomes challenging in the presence of noise, particularly under low SNR or under noise types, such as electrode movement artifacts, which overlap with the frequency range of the T-wave. These artifacts can change the T-wave shape and are not easily removed by simple frequency domain filtering.

\begin{figure}[tb]
%\centering
\includegraphics[width=8.cm]{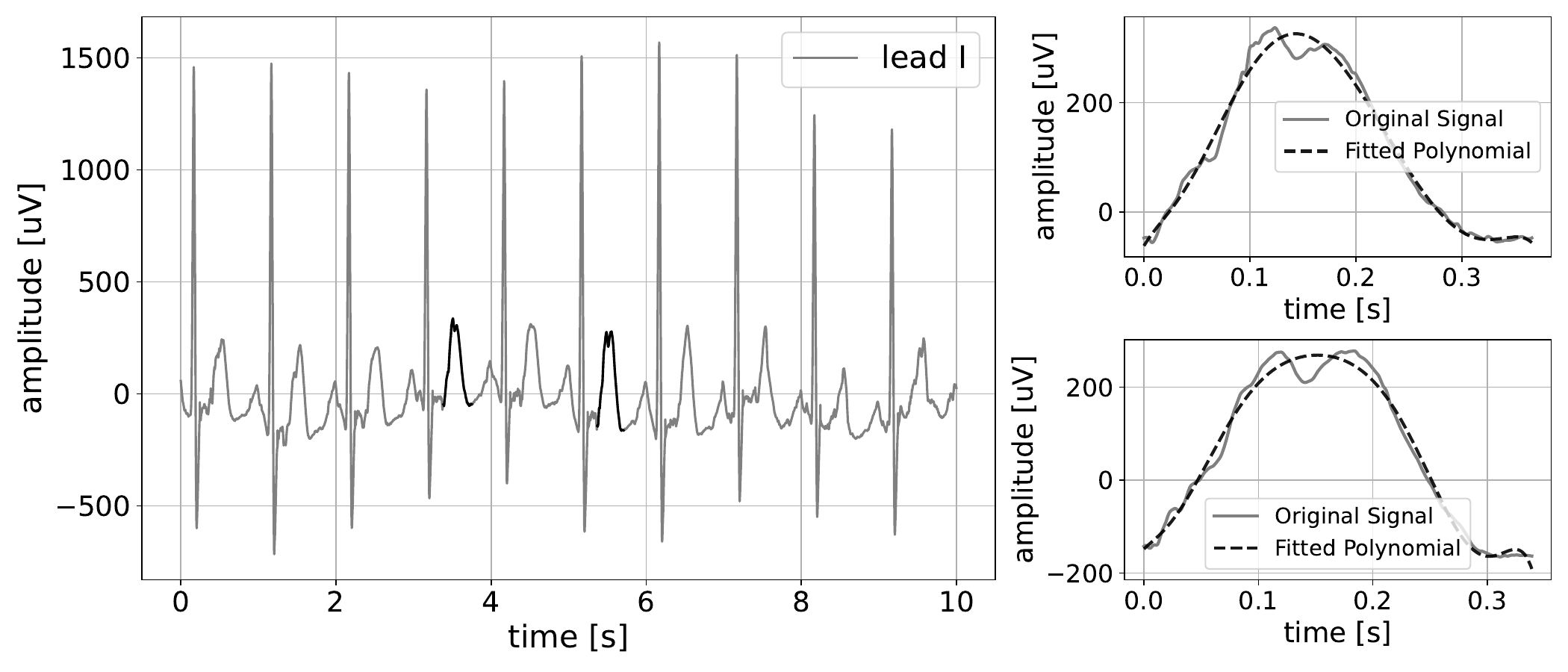}
\caption{\label{fig:polynomial_fit} Example of polynomial modeling of T-waves. On the left a recording corrupted by electrode movement noise with an SNR of 10\,dB is shown. The highlighted T-waves in this noisy signal are then modeled using an 8th-degree polynomial, as shown on the right.
}
\end{figure}

To address these challenges, we employed a model-based approach. An 8th-degree polynomial was fitted to the T-wave using least squares error model fitting. As shown in previous studies~\cite{Roonizi2013,Fattahi2022}, model-based ECG parameter estimation can improve the accuracy of amplitude estimation in the presence of noise. An example of this polynomial fitting is illustrated in Fig.~\ref{fig:polynomial_fit}, which depicts a signal exhibiting TWA at an amplitude of 47\,$\mu$V, a heart rate of 60 beats per minute, and a respiratory rate of 12 breaths per minute, contaminated with electrode movement noise at an SNR of 10\,dB. The right side of the figure displays the T-waves highlighted in the signal, showing their appearance before and after polynomial modeling.

After modeling the T-wave, we identified the maximum point of the fitted curve, based on the known concave shape of the T-wave in lead I (used to generate the synthetic ECG). This process resulted in a sequence of model-based extracted T-wave amplitudes.

To evaluate the performance of the T-wave amplitude detection, we employed the mean absolute error (MAE) as the primary metric. We explored distinct scenarios resulting in four datasets listed in Table~\ref{tab:dataset}:
\begin{itemize}
    \item \textit{Dataset 1:} Clean synthetic ECGs without artifacts.
    \item \textit{Dataset 2:} Synthetic ECGs with respiratory modulation.
    \item \textit{Dataset 3:} Synthetic ECGs without respiratory modulation but with varying types and levels of artifacts.
    \item \textit{Dataset 4:} Synthetic ECGs with respiratory modulation combined with varying types and levels of noise.
\end{itemize}
This evaluation was performed for both raw and model-fitted T-wave amplitudes.

\section{T-wave alternans level estimation}
After extracting sequences of raw and model-fitted T-wave amplitudes from each synthetic recording, we applied a Modified Moving Average (MMA)~\cite{nearing2002mma}, to estimate the level of TWA. We evaluated the MAE and standard deviation between the TWA estimates obtained using MMA and the ground truth TWA for dataset type (4), i.e., recordings with respiratory modulation plus noise corruption, across six different SNRs for each noise type. This analysis was performed for both raw and model-fitted T-waves, and the results were then compared.

\section{Detection of T-wave Alternans}
Complementing T-wave amplitude estimation, we also studied TWA detection on dataset type (4) (recordings with respiratory modulation plus noise corruption) as this most closely resembles real-world ECG recordings. All detection methods were applied to both sequences of raw and model-fitted T-wave amplitudes.

%%All detection methods were applied to both raw and modeled T-wave amplitudes. However, due to space limitations, we report only the detection results on the modeled T-waves, which demonstrated improved performance in T-wave amplitude estimation. 

%%All detection methods were applied to both raw and modeled T wave amplitudes. Due to the sapce limitation we will report only the results of detection on the modeled T-waves due to its improved performance in T-wave amplitude estimation as well as the prior expectation that modeling enhances amplitude estimation in noisy environments.

%%Due to the spaced limitation, all detection methods were applied to the amplitudes of the modeled T-waves. This choice was based on our findings, which demonstrate improved performance in T-wave amplitude estimation, as well as the prior expectation that modeling enhances amplitude estimation in noisy environments.

\subsection{Surrogate data analysis}
We combined the MMA method with surrogate data analysis (SDA)~\cite{nematitwa} to determine whether the detected T-wave alternans amplitudes are real or mere artifacts of noise. Presumably, if the temporal order of the T-wave amplitudes conveys information, disrupting the temporal order of beats by shuffling the beat sequence should significantly reduce the beat-to-beat alternation amplitude. This approach helps identify noise-induced alternating patterns (NIAP), defined as alternating patterns in beat sequences that arise from sources other than true alternation in ventricular repolarization. To estimate NIAP, we generated surrogate data by repeatedly reshuffling the beat sequence ($N = 250$) and calculating the alternans amplitude for each surrogate arrangement using MMA. We then conducted a statistical test by comparing the observed TWA amplitude to an upper percentile ((1 $-$ $\alpha$) × 100) of the NIAP distribution (here, the 95th percentile, with $\alpha$ = 0.05). If the measured TWA amplitude equals or exceeds the NIAP values up to the 95th percentile, the TWA amplitude was considered significant. Otherwise, it was deemed indeterminate for the analysis window \cite{nematitwa}.
For each recording, we determined whether the TWA was significant (classified as 1) or indeterminate (classified as 0), thereby distinguishing true alternans from those influenced by noise.

%%Additionally, we calculated the percentage of significant TWA detections based on SDA for each scenario. This analysis was conducted for both raw and modeled T-waves, and the results were subsequently compared.

\subsection{T-Wave amplitude alternation state transition matrix}
In addition to surrogate data analysis, we developed a classifier based on the probability of alternating T-waves. This involves constructing a state transition matrix (STM) using a first order Markov chain to model the sequence of T-wave amplitudes~\cite{Clifford2010}.

The process starts with a time sequence of T-wave amplitudes $t_k$ ($k = 1, \ldots, K$) extracted from an ECG. To detect TWA, the T-wave amplitudes are first converted into a binary high/low (up/down) sequence $a_k$:
\begin{equation}
    a_k = \left\{
         \begin{array}{ll}
         \text{High}, & t_k \geq t_{k-1} \\
         \text{Low}, & t_k < t_{k-1}
    \end{array}\right.
\end{equation}
% Specifically, an amplitude value is classified as high (1) if it is greater than or equal to the preceding value and as low (0) otherwise.
We then calculate the percentage of transitions in $a_k$ in four cases: low to low ($p_{\text{LL}}$), low to high ($p_{\text{LH}}$), high to low ($p_{\text{HL}}$), and high to high ($p_{\text{HH}}$). These percentages are used to form a state transition matrix (STM)~\cite{Clifford2010}:
\begin{equation}
    \text{STM} = \left(
    \begin{array}{cc}
    p_{\text{LL}}  & p_{\text{LH}} \\
    p_{\text{HL}}  & p_{\text{HH}} 
    \end{array}
    \right)
\label{eq:stm}    
\end{equation}
Apparently, the sum of each row of $\text{STM}$ is equal to 1 ($p_{\text{LL}}  + p_{\text{LH}} = 1$ and $p_{\text{HL}}  + p_{\text{HH}} = 1$). Therefore, for further analysis, we simply focus on the off-diagonal entries $p_{\text{LH}}$ and $p_{\text{HL}}$ to assess the alternations.

% The matrix summarizing these state transitions is presented in Table~\ref{tab:stm_table}.
% \begin{table}[tb]
%     \centering
%      \caption{State transition matrix}
%     \label{tab:stm_table}
%     \begin{tabular}{|c|c|}
%         \hline
%         High to High & High to Low\\ \hline
%         Low to High & Low to Low \\ \hline
%     \end{tabular}
% \end{table}

% Once we have the transition counts for each state, we convert them into estimates of high-low amplitude probabilities by normalizing the \texttt{stm}:

% \[
% \text{STM} = \text{diag}\left(\frac{1}{\sum_{i} \text{stm}_{i}}\right) \cdot \text{stm}
% \]

% This normalization process adjusts the raw counts to reflect the probabilities of transitioning between states, providing a more interpretable representation of the state dynamics.

% In this normalized STM, the sums of each row and column are equal to 1.

\subsubsection{STM Corner Cases}
\begin{figure}[tb]
\centering
\includegraphics[trim={0 0cm 0 1.2cm},clip,width=6.0cm]{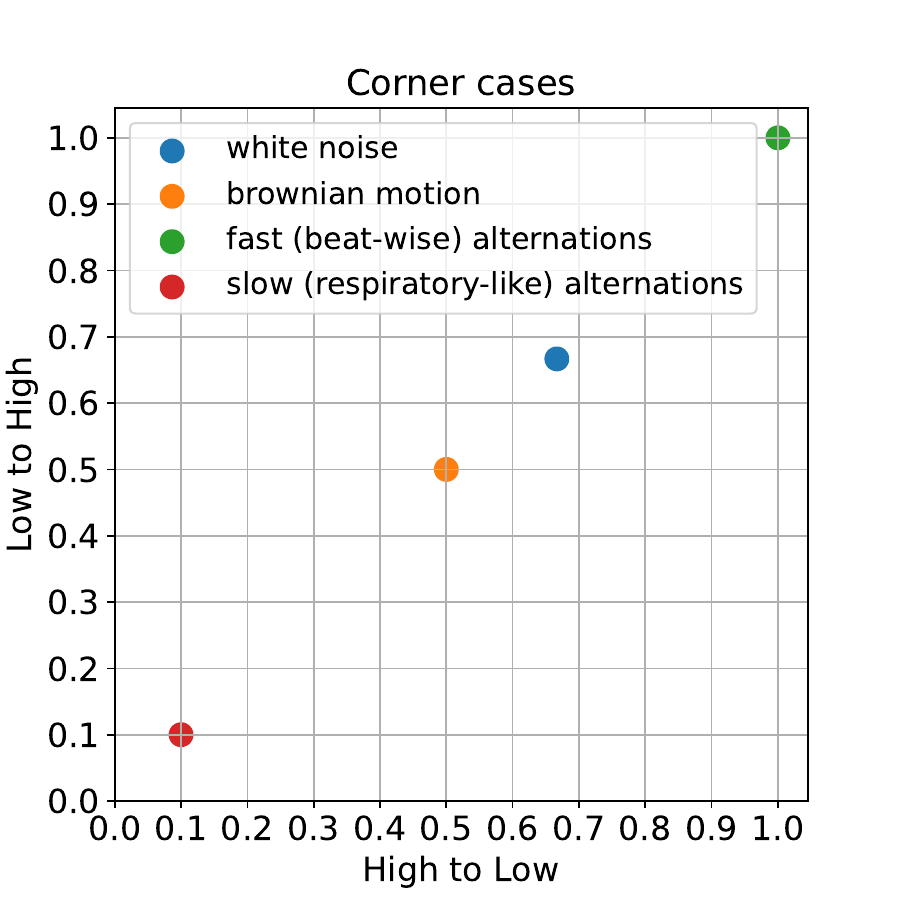}
\caption{Example of corner cases for off-diagonal probabilities from the STM defined in \eqref{eq:stm}, illustrating the expected STM patterns for different types of amplitude sequences: (green) perfect beat-wise T-wave alternans; (red) slowly varying periodic patterns such as respiration effects; (orange) Brownian motion (Wiener process), resulting in equal transition probabilities; and (blue) white noise, where frequent state switching leads to a higher probability of transitions than self-repetitions. These cases serve as reference cases for interpreting STMs derived from real or synthetic ECG.}
\label{fig:corner_cases}
\end{figure}

Several corner cases exist for the T-wave amplitude alternation STM, which are illustrated in Fig.~\ref{fig:corner_cases}. Accordingly, in the case of perfect TWA, where high and low T-wave amplitudes alternate strictly every other beat, the STM becomes:
\begin{equation*}
\text{STM}_{\text{TWA}} = \begin{pmatrix}
0 & 1 \\
1 & 0 
\end{pmatrix}
\end{equation*}
This indicates that every low is followed by a high amplitude, and every high is followed by a low amplitude---representing perfect alternans.
For slowly varying periodic amplitude sequences $t_k$, such as those caused by respiration, the STM becomes nearly diagonal:
\begin{equation*}
\text{STM}_{\text{slow}} = \begin{pmatrix}
1 - \epsilon & \epsilon \\
\epsilon & 1 - \epsilon
\end{pmatrix}
\end{equation*}
where $\epsilon$ is a small value, showing that transitions between states are much less frequent than the number of ECG beats.
When the amplitude sequence $t_k$ is a Wiener process (Brownian motion), $a_k$ is White noise and the high-low transitions are equally likely in both directions:
\begin{equation*}
\text{STM}_{\text{Brownian}} = \begin{pmatrix}
0.5 & 0.5 \\
0.5 & 0.5
\end{pmatrix}
\end{equation*}
Finally, when $t_k$ is white noise, due to the uncorrelatedness of $t_k$ and $t_{k-1}$, it results in more frequent switching between states, which can be shown to converge to:
\begin{equation*}
\text{STM}_{\text{white}} = \displaystyle\begin{pmatrix}
\frac{1}{3} & \frac{2}{3} \\[1ex]
\frac{2}{3} & \frac{1}{3}
\end{pmatrix}
\end{equation*}
These corner cases help interpret STM values from real or synthetic ECG.

\subsection{T-wave alternans classification}
\begin{figure}[tb]
\centering
\includegraphics[width=1.0\columnwidth]{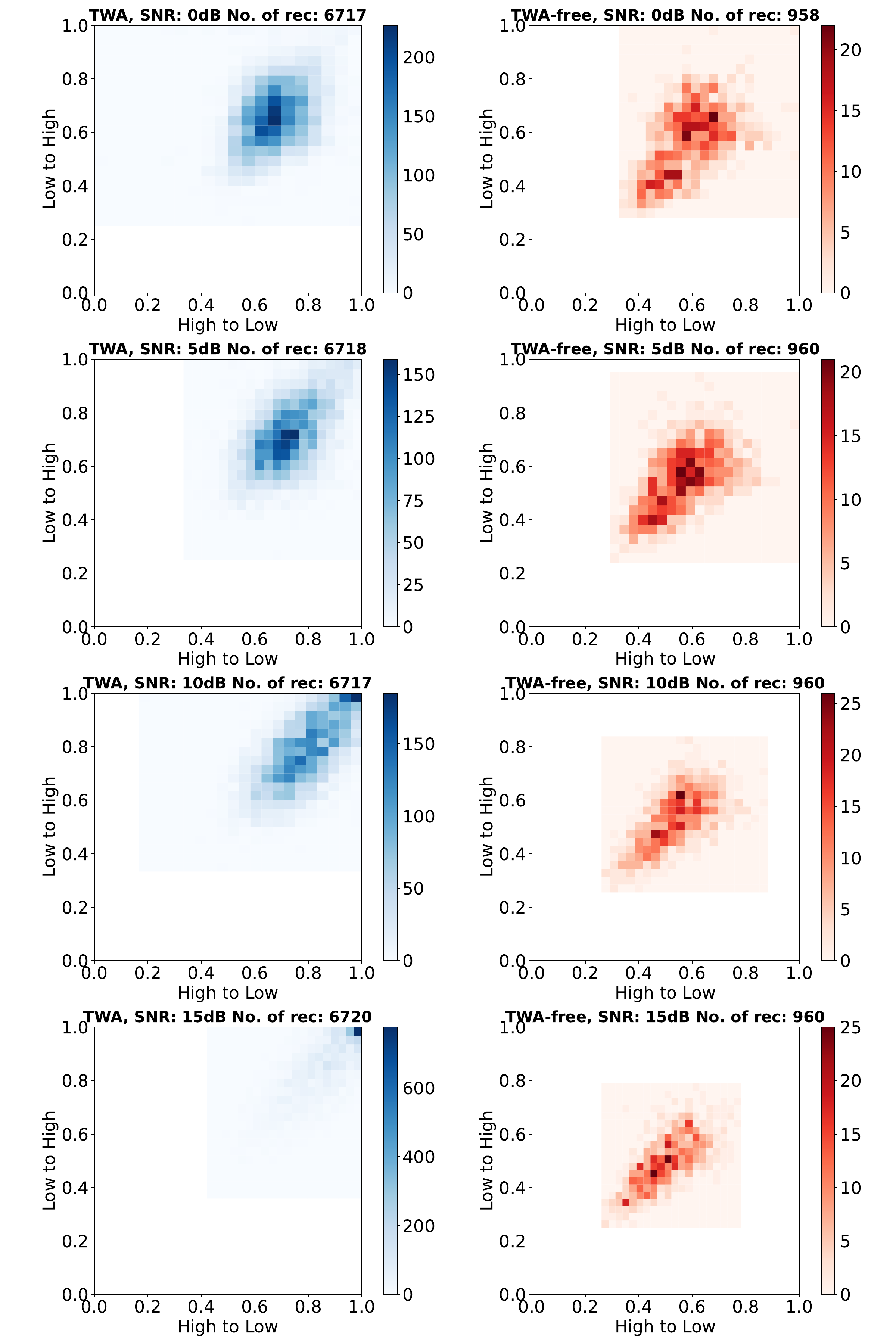}
\caption{Distribution heatmaps of STM low-to-high ($p_{\text{LH}}$) and high-to-low ($p_{\text{HL}}$) transition probabilities for synthetic ECG with and without TWA in different SNR levels (0, 5, 10, and 15\,dB), per row. In TWA-present cases, both off-diagonal probabilities converge to 1 as SNR increases, indicating increasingly reliable detection of TWA. TWA-free cases exhibit more dispersed distributions centered around 0.3 to 0.7, reflecting randomness in the absence of true TWA.}
\label{fig:heatmaps_twa}
\end{figure}

Fig.~\ref{fig:heatmaps_twa} shows the distributions of low-to-high ($p_{\text{LH}}$) and high-to-low ($p_{\text{HL}}$) probabilities for synthetic signals with and without TWA, across four SNR levels (0 to 15\,dB). In signals with TWA, as SNR increases, both off-diagonal probabilities converge to one, indicating strong alternation. In contrast, signals without TWA show broader distributions centered around 0.3 to 0.7, reflecting randomness or weak structure.

% From the studied corner cases, it is evident that for perfect periodic T-wave alternans the corresponding probabilities in the off-diagonal entries of the STM are equal to 1. Conversely, for a slowly varying periodic process, such as one resembling respiration, the probabilities approach 0. In the case of Brownian motion, the STM probabilities are 0.5, while for white noise, they are approximately 0.7.

% Fig. \ref{tab:heatmaps_twa} captures distribution of low to high and high to low probabilities for our synthetic data with and without TWA. Four rows represents different levels of SNR spanning from 0 to 15 dB. For data containing TWA, the histograms demonstrate that as signal quality improves (higher SNR), the low-to-high and high-to-low probabilities converge toward one, indicating increased confidence in alternations. Conversely, in data without TWA, the probabilities cluster toward mid-range values (approximately 0.3 to 0.7), reflecting the absence of TWA patterns.

%%\subsection{T-wave alternans classification}
To distinguish between recordings containing TWA and those without we developed a logistic regression classifier using low-high (LH) and high-low (HL) transitions from STM as the model input. The dataset was divided into training and validation sets using a stratified sampling approach to ensure a similar distribution of classes across both sets.

To determine the optimal threshold for the posterior probability from the logistic regression model, we used the operating point of the receiver operating characteristic (ROC) curve, estimated at 0.85 on the validation set. 
To evaluate the performance of the SDA and STM method, we employed classification metrics including sensitivity, positive predictive value (PPV), F1-score, and specificity. Results for both detection methods are reported on the validation set.

We also explored the impact of the number of beats included in the analysis. Generally, TWA identification can be explored from an estimation and detection theoretical perspective~\cite{van2004detection}, where both SNR and the number of beats are critical factors influencing the detection threshold, false alarm rate, and overall sensitivity. As the number of beats increases, statistical confidence in distinguishing structured alternans from noise improves, potentially lowering the minimum detectable alternans amplitude under a given noise level. 

To address this, we calculated the F1-score for both the SDA and STM methods, varying the SNR from -5 to 30\,dB and the number of beats from 10 to 60, with a step size of 10 beats. This approach allows us to assess how the number of beats used in averaging influences detection performance, as a higher number of beats improves the SNR and the ability to reliably detect TWA.

%%%This analysis aligns with the theoretical framework of Shannon’s Information Theory \cite{shannon1948mathematical}, where both SNR and the number of beats are critical factors that influence the detection threshold and performance.

% 3x3 grid 3SNRx3types of artifacts

%%\section{TWA distribution}
%%\textcolor{red}{IDEA (Reza): Add a note on why a ``folder gaussian distribution'' is a good replacement for the Gamma distribution in this paper: \url{https://lcp.mit.edu/pdf/NematiTBE10.pdf}. This will give us a model-based approach to provide statistical power on whether or not a specific TWA value is statistically relevant.
%%}

\section{Results}
\subsection{T-wave amplitude analysis}
To evaluate the accuracy of T-wave amplitude measurement, we first calculated the MAE between the ground truth T-wave amplitudes and the detected T-wave amplitudes for clean ECG signals (Dataset 1). Using the raw T-wave amplitude sequence, we obtained an MAE of 1.28 \textpm 2.9\,$\mu$V, while the modeled T-wave amplitude sequence resulted in an MAE of 3.28 \textpm 3.26\,$\mu$V.

Upon introducing respiratory modulation (Dataset 2), the MAE increased notably, with values of 19.9 \textpm 23.02\,$\mu$V for raw T-wave amplitude detection and 19.8 \textpm 22.92\,$\mu$V for modeled T-wave amplitude detection.

\begin{figure}[tb]
\centering
\begin{subfigure}{\columnwidth}
  \centering
  \includegraphics[trim={0 .8cm 0 1cm},clip,width=\linewidth]{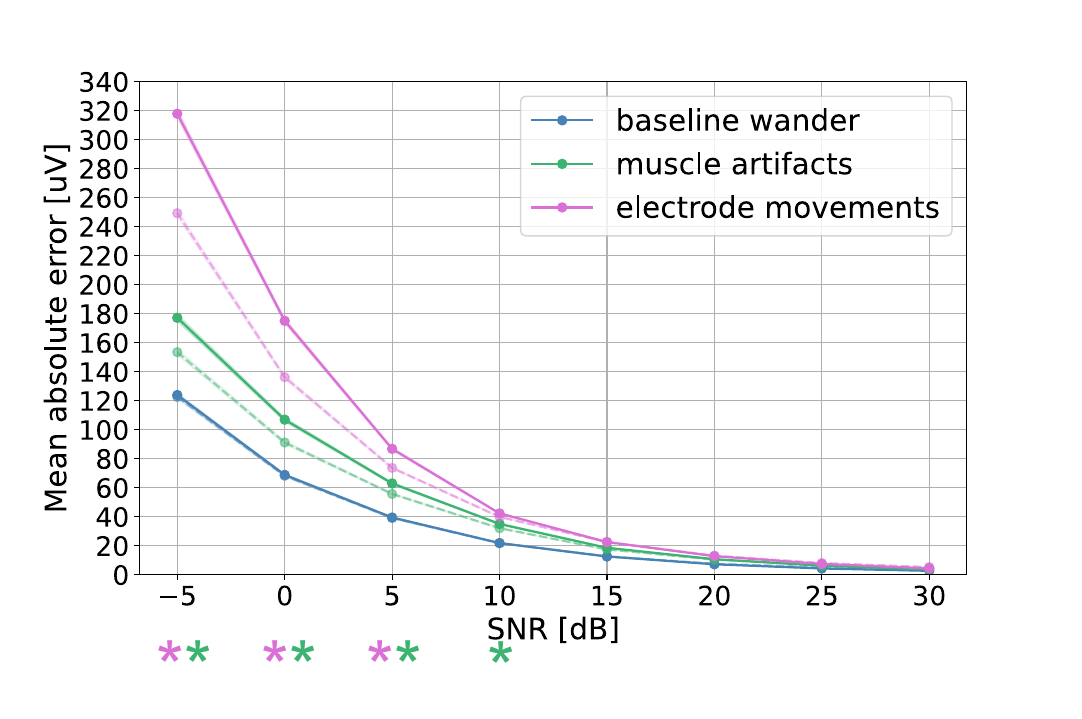}
  \caption{Without respiratory modulation}
  \label{fig:mae_noise_without_resp}
\end{subfigure}
% \vspace{1em}
\begin{subfigure}{\columnwidth}
  \centering
  \includegraphics[trim={0 .8cm 0 1cm},clip,width=\linewidth]{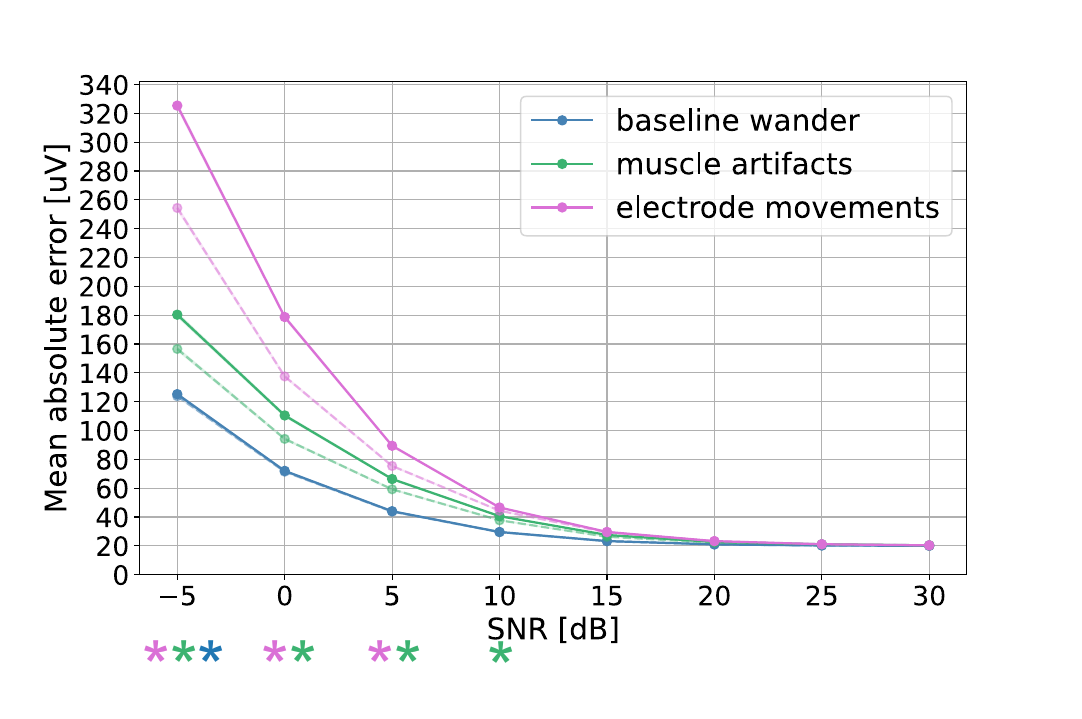}
  \caption{With respiratory modulation}
  \label{fig:mae_noise_with_resp}
\end{subfigure}
\caption{Results of T-wave amplitude measurement with added noise, without (a) and with (b) respiratory modulation. Mean absolute error (MAE) with 95\% confidence intervals was computed between the ground truth and both raw and modeled T-wave amplitudes (modeling results shown as dashed lines), across different noise types and SNR levels. Stars (color-coded by noise type) indicate cases with statistically significant improvements ($p<0.05$) of the modeled over the raw approach.}
\label{fig:mae_noise_combined}
\end{figure}

Next, we examined the scenario without respiratory modulation but with the addition of three types of noise at different SNR levels (Dataset 3). As detailed in Section~\ref{sec:noise_gen}, the noises include baseline wander, muscle artifacts, and electrode movement from the MIT-BIH Noise Stress Test Database~\cite{moody1984noise,goldberger2000physiobank}, which we resampled to the ECG sampling frequency, scaled, and added to the synthetic ECG at different proportions. We observed that at the highest SNR, the error approached 3\,$\mu$V for both the raw and modeling approaches for all three noise types. These results are presented in Fig.~\ref{fig:mae_noise_without_resp}. 

Finally, we examined the impact of respiratory modulation combined with different noise types on T-wave measurement performance across different SNR levels (Dataset 4), which are presented in Fig.~\ref{fig:mae_noise_with_resp}. A significant improvement using the modeling was observed, particularly with electrode movement and muscle artifact noise at -5, 0, 5, and 10\,dB, as confirmed by the Mann–Whitney U test with a significance threshold set at 0.05 (Fig.~\ref{fig:mae_noise_with_resp}). Overall, the MAE decreased with increasing SNR for both raw and modeled T-wave amplitudes.

%%at SNRs of 5 and 10 dB, where polynomial modeling reduced the MAE from 89.39 µV to 75.38 µV and from 46.57 µV to 44.34 µV, respectively. Similarly, for muscle artifact noise at SNRs of 5 and 10 dB, the MAE decreased from 66.31 µV to 59.17 µV and from 40.54 µV to 37.69 µV.

%%The reduction in absolute error for muscle artifacts noise at 5 
%%and 10 dB  

%%was statistically significant, as confirmed by the Mann-Whitney U test with a significance threshold set at 0.05. However, the reduction in absolute error for electrode movements noise was significant only at 5 dB level of noise.
%(($p = 2.25\mathrm{e}{-91}$). 

%%The relationship between MAE and SNR across different noise conditions with respiratory modulated signal is illustrated in Fig. \ref{fig:mae_noise_without_resp} B.

\subsection{T-wave alternans level estimation}
%%In this method, we used only data containing TWA values of 30 µV or higher. 

\begin{figure}[tb]
%\centering
\includegraphics[trim={0 .5cm 0 1.2cm},clip,width=8.3cm]{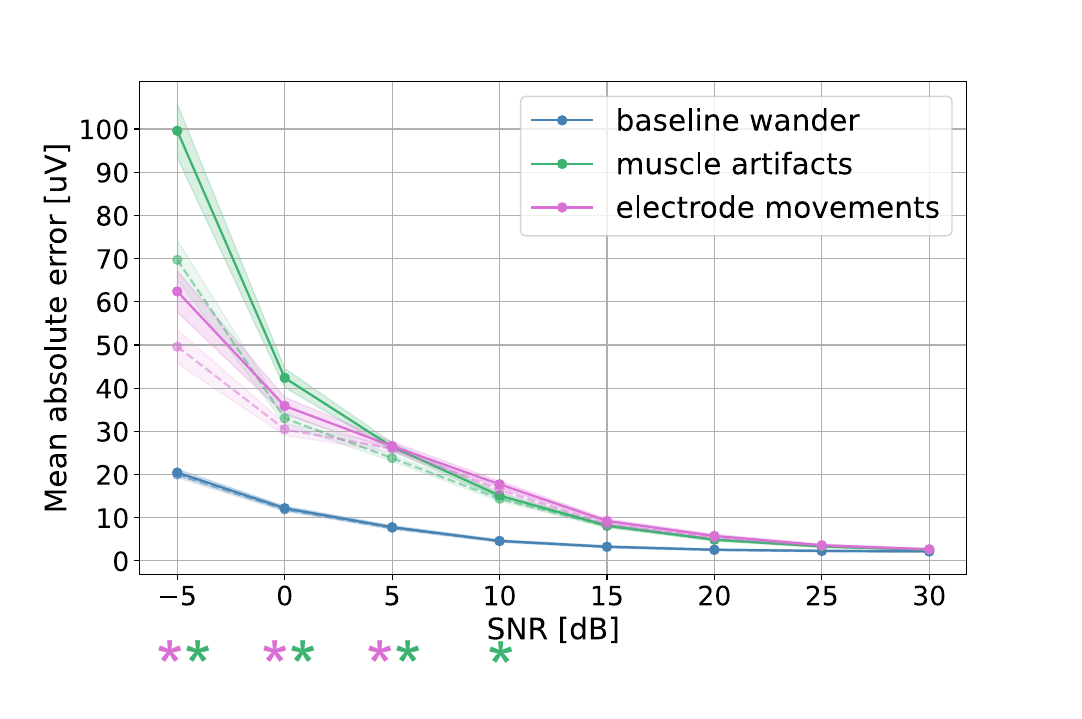}
\caption{Results of T-wave Alternans estimation using the Modified Moving Average (MMA) method on recordings with respiratory modulation and added noise. Mean absolute error (MAE) with 95\% confidence intervals was computed between the ground truth TWA amplitude and both raw and modeled TWA estimates (modeling results shown as dashed lines), across various noise types and SNR levels. Color-coded stars indicate cases where the modeled approach significantly outperformed the raw approach ($p < 0.05$).
}
\label{fig:mae_mma_result}
\end{figure}

To assess the accuracy of TWA estimation using the MMA method we calculated the MAE between the estimated TWA using the MMA method and the ground truth TWA. The relationship between SNR and MAE with 95\% condifedence intervals is shown in Fig.~\ref{fig:mae_mma_result}. The most notable difference between the raw and modeling approaches was observed in low SNRs. Significant improvements were achieved for MA and EM noise at -5, 0, 5 and 10\,dB using the Mann-Whitney U test with a significance threshold set at 0.05. %at 10 (from 17.74 to 16.46 uV) and 15dB (from 9.24 to 8.48 uV) and for EM noise at 5 dB (from 26.43 to 23.8 uV) at significance level of p \textless 0.05.

%%surrogate data analysis, where we calculated the percentage of recordings with significant TWA. For both electrode movement and muscle artifact noise at lower SNR levels, we observed an improvement with the modeling approach. The percentage of recordings with containing significant TWA according to the SDA is presented in Table \ref{tab:sda}. No improvement was observed for baseline wander, likely due to the distinct nature of noise which is easily suppressed using the high-pass filter.

%%\begin{table}[tb]
%%\centering
%%\caption{Results of surrogate data analysis 
%%\\Percentage of  recordings containing significant T-wave alternans}
%%\label{tab:sda}
%%\begin{tabular}{|c|c|c|c|c|}
%%\hline
%%& \multicolumn{2}{c|}{MA} & \multicolumn{2}{c|}{EM} \\
%%\hline
%%SNR (dB) & Raw (\%) & Modeled (\%) & Raw (\%) & Modeled (\%) \\
%\hline
%%5 & 41.83 & 47.41 & 44.96 & 49.38 \\
%%\hline
%%10 & 78.11 & 81.9 & 81.65 & 83.75 \\
%%\hline
%%15 & 94.46 & 94.96 & 96.29 & 96.65 \\
%%\hline
%%\end{tabular}
%%\end{table}

\subsection{Detection T-wave Alternans}
We assessed the binary classification of records into TWA-free and TWA-present categories using the surrogate data analysis (SDA) and the state transition matrix (STM) methods using logistic regression. These methods were applied exclusively to the most realistic dataset, which included recordings corrupted by both respiratory modulation and additive noise (Dataset 4). Results for TWA detection focuses on raw T-waves, not the modeled T-waves.
%%%, as modeling improved TWA estimation but did not enhance TWA detection. 

\begin{figure}[tb]
%\centering
\includegraphics[trim={0 0cm 0 .6cm},clip,width=9cm]{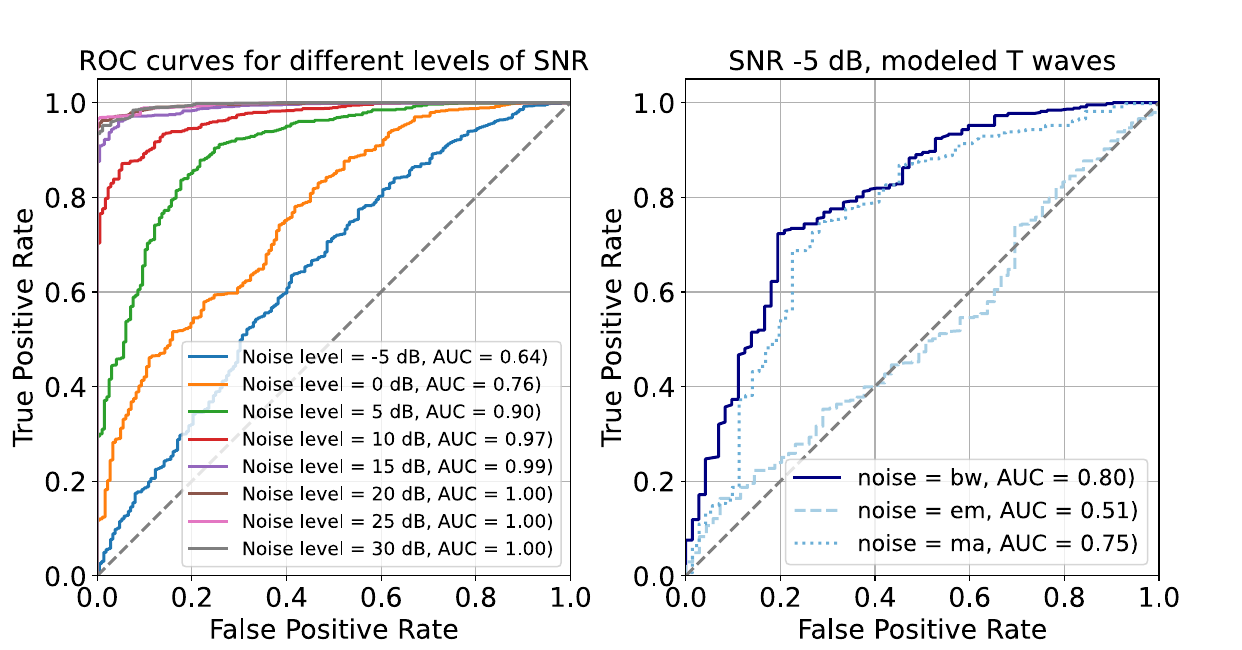}
\caption{\label{fig:rocs}Receiver Operating Characteristic curves derived from logistic regression results across varying SNRs. Additionally, for the lowest SNR of -5\,dB, separate ROC curves are presented for different noise types, including muscle artifacts (ma), electrode movements (em), and baseline wander (bw).}
\end{figure}

Initially, we evaluated the STM method based on the ROC curves at various noise levels, as shown in Fig.~\ref{fig:rocs}. These results show that higher SNRs correspond to strictly improved the ROC curves and increased AUROC values~\cite{Sameni2025ROC}, as depicted in the figure. Further analysis at a low SNR level (-5 dB) shows significant differences in AUROC depending on the noise type (Fig.~\ref{fig:rocs}). Baseline wander, which has lower spectral overlap with the ECG and can be effectively mitigated with high-pass filtering, achieved an AUROC of 0.8 at -5\,dB. Muscle artifacts resulted in an AUROC of 0.75, while electrode movement noise, which significantly overlap with the ECG spectra, demonstrated the lowest performance, with an AUROC of 0.51.

Table~\ref{tab:sda_stm_voting} summarizes the results of all three detection approaches in terms of sensitivity, specificity, positive predictive value (PPV) and F1-score.
\begin{table}[tb]
\centering
\caption{Results of surrogate data analysis and state transition matrix}
\begin{tabular}{c|c|c}
\hline
\textbf{Metric} & \textbf{STM} & \textbf{SDA} \\
\hline
Sensitivity & 0.87 & 0.70  \\
\hline
PPV & 0.98 & 0.99  \\
\hline
%%NPV & 0.61 & 0.64 & 0.88 \\
%%\hline
Specificity & 0.87 & 0.98  \\
\hline
F1-score & 0.92 & 0.82  \\
\hline
\end{tabular}
\label{tab:sda_stm_voting}
\end{table}

\begin{figure}[tb]
\centering
\includegraphics[width=1.0\columnwidth]{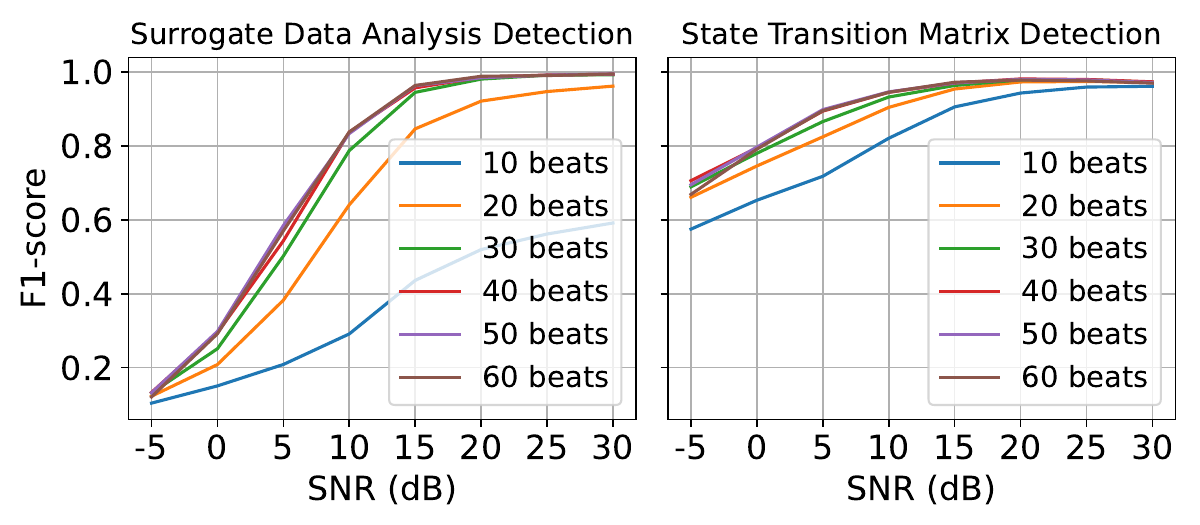}
\caption{\label{fig:num_beats_impact} F1 scores for both TWA detection methods, with varying numbers of analyzed beats at different SNR. TWA detection improves with the number of ECG beats.}
\end{figure}

Framing the TWA detection problem within the context of detection and estimation theory~\cite{van2004detection}, we recognize that performance is influenced not only by the noise level and TWA amplitude but also by the number of beats included in the analysis. Fig.~\ref{fig:num_beats_impact} illustrates how detection performance, in terms of F1-score, changes with the increasing number of beats, alongside varying SNR levels, for both SDA and STM methods, as the number of beats is varied from 10 to 60. As expected, TWA detection performance improves for both methods with an increased number of beats. This reflects results from information theory and communication theory that longer or more redundant signals offer increased robustness to noise~\cite{shanmugan1979digital,cover2006elements,van2004detection}.

%%\begin{figure}[tb]
%\centering
%%\includegraphics[width=8.3cm]{images/f1_score.png}
%%\caption{\label{tab:f1_score}F1-scores for T-wave Alternans detection at different SNR levels using Surrogate Data Analysis (SDA) and State Transition Matrix (STM) on recordings with respiratory modulation and noise. }
%%Dashed lines indicate the modeling approach, while solid lines represent results from raw T-waves.}
%%\end{figure}

%%\begin{table}[tb]
%%\centering
%%\caption{Results of surrogate data analysis, state transition matrix classification, and combination of the two for noise level of 5dB}
%%\label{tab:sda_stm_voting_5}
%%\begin{tabular}{|c|c|c|c|}
%%\hline
%%\multicolumn{4}{|c|}{A. Results for Raw T-wave ammplitudes} \\
%%\hline
%%Metric & SDA & STM & Combined (OR condition) \\
%%\hline
%%Sensitivity & 0.60 & 0.82 & 0.90 \\
%%\hline
%%PPV & 0.97 & 0.97 & 0.96 \\
%%\hline
%%NPV & 0.25 & 0.40 & 0.53 \\
%%\hline
%%Specificity & 0.88 & 0.80 & 0.72 \\
%%\hline
%%F1-score & 0.74 & 0.88 & 0.93 \\
%%\hline
%%\multicolumn{4}{|c|}{B. Results for Modeled T-Wave Amplitudes} \\
%%\hline
%%Metric & SDA & STM & Combined (OR condition) \\
%%\hline
%%Sensitivity & 0.65 & 0.79 & 0.91 \\
%%\hline
%%PPV & 0.97 & 0.98 & 0.96 \\
%%\hline
%%NPV & 0.27 & 0.38 & 0.55 \\
%%\hline
%%Specificity & 0.88 & 0.87 & 0.78 \\
%%\hline
%%F1-score & 0.78 & 0.87 & 0.93 \\
%%\hline
%%\end{tabular}
%%\end{table}

\section{Discussion}
\label{sec:discussion}
Despite the well-documented significance of TWA in cardiovascular risk assessment, the lack of calibrated and annotated reference datasets necessitated the use of a synthetic ECG model~\cite{McSharry2003,SCJS06}, which generates realistic waveforms capable of passing the so-called Turing test of realism.
%%\textcolor{red}{Reza - Add a table to summarize the takeaways of this study: noise type, noise level, respiration, averaging window length, modeling vs raw data, etc.}

%%\textcolor{blue}{Reza to add a discussion on the TWA detection problem being a standard detection problem with theoretical thresholds of performance for detecting (bits) following Shannon's Information Theory. We can show that both the amplitude and signal length (number of beats used to average) plays a role}

Existing methods for TWA estimation have been largely evaluated on noise-free recordings, as recordings with low signal quality are typically deemed unusable for analysis. The current guidelines for TWA measurements using the MMA method rely on symptom-limited exercise stress testing, during which patients retain chronic medications. However, physical movements during these tests can cause electrode disturbances and muscle artifacts, resulting in poor-quality signals that are often discarded from analysis. When transitioning TWA estimation and detection to Holter monitor data, the data volume increases significantly, but the overall signal quality tends to decrease as data is collected in daily environments. Discarding all low-quality recordings in such settings would result in substantial data loss. Our model-based TWA estimation and detection algorithm addresses this need and minimizes the impact of noise-induced TWA.

The first contribution of this research was T-wave modeling, aimed at improving the accuracy of measured T-wave amplitudes under variable noise levels. As shown in Fig.~\ref{fig:mae_noise_without_resp}, when modeling was applied, T-wave amplitude estimation performance improved significantly for both MA and EM noise at low SNRs (-5, 0, 5 and 10\,dB).
Additionally, the modeling approach enhanced TWA level estimation using the MMA method at low SNRs, as demonstrated in Fig.~\ref{fig:mae_mma_result}. For MA noise and EM noise at -5, 0, 5 and 10\,dB, modeling resulted in notable improvements. 

As expected, for both T-wave amplitude measurement and TWA estimation, MAE decreases as SNR increases. The impact of modeling on TWA estimation performance diminishes at higher SNRs since the signal quality is already sufficient, and polynomial modeling does not further improve amplitude measurements. In this study, we employed polynomial modeling with a degree of 8. Empirically, the 8th-degree polynomial provided better performance for T-wave amplitude detection compared to the 6th or 10th degrees, and was therefore selected for this analysis. Apparently, the accuracy of model-based results is inherently constrained by the model's order. Variations in the polynomial degree or model type (e.g. sum-of-gaussian kernels) may produce different outcomes~\cite{Fattahi2022,Roonizi2013}.

It is important to highlight the impact of respiration on T-wave amplitude measurement, which has been overlooked in the TWA literature and according to our results often leads to an overestimation of TWA. When measuring T-wave amplitude with only additive noise (Fig.~\ref{fig:mae_noise_without_resp}), the MAE in a high-quality signal of 30\,dB SNR was approximately 3$\mu$V across all noise types. However, when incorporating both respiratory modulation and additive noise, the MAE in 30\,dB SNR increases to 19$\mu$V (Fig.~\ref{fig:mae_noise_with_resp}), demonstrating the impact of respiration on T-wave amplitude measurement.
We acknowledge that the MAE of 19$\mu$V is also impacted by the chosen respiratory amplitude of 10\% amplitude variation, which we found to be representative of real ECG recordings. Adjusting this parameter could yield different MAE values. Nevertheless, the resulting errors seem to diminish when estimating TWA levels using the MMA (Fig.~\ref{fig:mae_mma_result}).

The lower the SNR, the less precise the TWA estimation becomes, even with T-wave modeling, which can lead to either overestimation or underestimation of TWA levels. To address this, we introduced a rigorous TWA detection using state transition matrix of low and high T-wave amplitude transitions. This technique was shown to effectively mitigate noise-induced TWA. From the ROC curve analysis of the STM method at a very low SNR of -5\,dB (Fig.~\ref{fig:rocs}), baseline wander (BW) and muscle artifact (MA) noise exhibit relatively strong performance, with AUROC values of 0.8 and 0.75, respectively. However, electrode movement (EM) noise performs poorly, with an AUROC of 0.51. This can be associated to the fact the EM noise has maximal overlap with the ECG spectra. EM noise is particularly prevalent under standard TWA collection conditions, where higher heart rates and physical activity often cause electrode disturbances, such as lead loosening, displacement, or cable movements. Thus, the -5\,dB to 0\,dB SNR range represents critical cases for EM noise, and our analysis indicates that TWA estimations in such conditions cannot be reliable.

A closer examination of TWA detection (Fig.~\ref{fig:num_beats_impact}) reveals that the STM method performs better for low-quality signals, while the SDA method is more effective for higher-quality signals (above 15 dB). Therefore, combining the more `specific' SDA method with the more `sensitive' STM method, as shown in Table~\ref{tab:sda_stm_voting}, using a rule-based decision mechanism could improve TWA detection performance across all SNR levels. Overall, the STM method achieves an F1-score of 0.92, compared to 0.81 for SDA, demonstrating a notable performance improvement. However, it is important to recognize that SDA was designed primarily to suppress the influence of TWA in noisy ECG recordings, where TWA detection may be unreliable. In contrast, STM was specifically developed to detect TWA, even under noisy conditions—explaining its superior performance.

%%As expected, detection performance improves with increasing signal quality.
The detection of TWA can be framed as a standard detection-estimation problem (similar to detecting sequence of bits in a digital communication system), where theoretical performance limits are determined by Shannon's Information Theory~\cite{cover2006elements}. Accordingly, the ability to detect TWA depends on the amplitude of alternations, noise level, noise type and the number of beats used for TWA assessment. A higher T-wave amplitude provides a stronger signal relative to background noise (higher SNR), improving detectability. Similarly, increasing the number of beats used in the averaging process enhances the detection SNR, reducing random variability and allowing more reliable discrimination of alternating patterns. These factors align with fundamental limits in detection theory, where information content (TWA level plus number of beats) and noise levels jointly determine the required SNR level for accurate detection~\cite{shanmugan1979digital,van2004detection,cover2006elements}. From this perspective, the results in Fig.~\ref{fig:num_beats_impact} highlight that as the SNR improves (with higher numbers of beats and better signal quality), TWA detection performance increases, which is consistent with Shannon's information theory. Understanding these constraints can help optimize detection methods and guide the design of more robust algorithms for TWA analysis.

The exact electrophysiological mechanism underlying TWA is not fully known. We acknowledge that our current approach, representing TWA as a simple bi-state alternation pattern, is a simplified approximation of real-world conditions. In practice, there are often more than two alternating beat types. In future work, we plan to extend the Markov model to higher orders, enabling the generation of synthetic ECGs with more complex rhythms and a wider variety of beat types.

Other modeling approaches, such as the sum-of-Gaussian kernels~\cite{McSharry2003,Fattahi2022,Roonizi2013}, may be more effective for TWA modeling, although they are computationally more demanding, making it more intensive to run on a beat-wise level across long datasets.

The Shannon Information theoretical perspective further suggests that a continuum of performance improvement is expected as the TWA amplitude and signal SNR improve. Here in we only explored non-alternating vs TWA at 30\,$\mu$V or higher. In future studies, a sensitivity analysis may be further performed to assess the performance of both T-wave estimation and TWA detection as the TWA level is varied more continuously from 0 to 30\,$\mu$V and beyond.

\section{Conclusion}
\label{sec:conclusion}
% In this study, we evaluated T-wave alternans estimation and detection under varying noise conditions and demonstrated that both tasks improve with increasing signal-to-noise ratios. Our results highlight the efficacy of polynomial modeling in enhancing T-wave amplitude estimation and reducing errors caused by muscle artifacts and electrode movements at low SNRs. We also showed that using detection methods, such as surrogate data analysis and state transition matrix approach could help us in detecting noise induced TWA and enable us to measure TWA even  in noisy environments. However, the performance of the methods remains limited by the presence of electrode movement noise, especially at low SNRs, which is common in and Holter monitoring scenarios.

% This work was conducted using synthetic ECG data, and while the findings provide valuable insights into TWA analysis in noisy settings, further validation on real-world data is required to confirm the clinical applicability and robustness of the proposed approaches.

In this study, we evaluated model-based methods for robust T-wave amplitude measurement and T-wave alternans (TWA) detection across varying noise conditions. Using a controlled synthetic ECG framework, we demonstrated that polynomial model-fitting prior to TWA detection significantly improves T-wave amplitude estimation under low SNRs, particularly in the presence of muscle artifacts and electrode movement noise. Additionally, we introduced a state transition matrix method for TWA detection, which, along with surrogate data analysis, enables more reliable discrimination of true TWA from noise-induced alternations.

Our results demonstrate that both estimation and detection performance benefit from higher SNRs and longer analysis windows (more ECG beats), consistent with theoretical expectations from information and detection theory. However, detection performance remains challenging under severe electrode movement artifacts, especially at low SNRs.

These findings highlight the potential of model-based approaches to improve TWA analysis in noisy environments, such as ambulatory Holter monitoring. Future work will involve extending the models to handle more complex alternation patterns and validating the proposed framework on real-world clinical data for specific outcomes.

% \textcolor{red}{EDITED UP TO THIS POINT - REZA}

\section*{References}
\bibliographystyle{IEEEtran}
\bibliography{References}
\end{document}